\documentclass{ws-jai}
\usepackage[flushleft]{threeparttable}
\newcommand\degr{\hbox{$^\circ$}}

\newcommand\arcmin{\hbox{$^\prime$}}
\newcommand\farcs{\hbox{$.\!\!^{\prime\prime}$}}
\newcommand\mnras{MNRAS}             % Monthly Notices of the Royal Astronomical Society
\newcommand\pasp{PASP}               % Publications of the Astronomical Society of the Pacific
\newcommand\aap{A\&A}                % Astronomy and Astrophysics
\newcommand\aj{AJ}                   % Astronomical Journal (the)
\newcommand\jai{JAI}                   % ournal of Astronomical Instrumentation
\newcommand\japa{JApA}                   %  Journal of Astrophysics and Astronomy
\newcommand\bsrsl{BSRSL}                   %  Bulletin de la Société Royale des Sciences de Liège
\newcommand\arXiv{arXiv}                   %  Bulletin de la Société Royale des Sciences de Liège
\newcommand\basi{BASI}                   %  Bulletin de la Société Royale des Sciences de Liège
\usepackage{graphicx}   % Including figure files
\usepackage{xcolor}

\begin{document}

\catchline{}{}{}{}{} % Publisher's Area please ignore

\markboth{Saurabh Sharma}{First LO Results with TIRCAM2}

\title{First Lunar Occultation Results with the TIRCAM2 
\\
Near-Infrared Imager at  the Devasthal 3.6-m Telescope
}

\author{Saurabh Sharma$^{1}$, Andrea Richichi$^{2}$, Devendra K. Ojha$^{3}$, Brajesh Kumar$^{1}$, Milind Naik$^{3}$, Jeewan Rawat$^{1}$,  
Darshan S. Bora$^{1}$,  Kuldeep Belwal$^{1}$, Prakash Dhami$^{1}$, and Mohit Bisht$^{1}$}

\address{
$^{1}$Aryabhatta Research Institute of Observational Sciences, Manora Peak, Nainital 263002, India, saurabh@aries.res.in\\
$^{2}$INAF – Osservatorio Astrofisico di Arcetri, Largo E. Fermi 5, I-50125 Firenze, Italy\\
$^{3}$Tata Institute of Fundamental Research, Homi Bhabha Road, Colaba, Mumbai 400005, India
}

\maketitle

\corres{$^{1}$Saurabh Sharma}

\begin{history}
\received{(to be inserted by publisher)};
\revised{(to be inserted by publisher)};
\accepted{(to be inserted by publisher)};
\end{history}

\begin{abstract}
TIRCAM2 is the facility near-infrared Imager at the Devasthal 3.6-m telescope in northern India, equipped with an Aladdin III InSb array detector.  We have pioneered the use of TIRCAM2 for very fast photometry, with the aim of recording Lunar Occultations (LO). This mode is now operational and publicly offered.  \\ In this  paper we describe the relevant instrumental details, we provide references to the LO method and the underlying data analysis procedures, and we list the LO events recorded so far.  Among the results, we highlight a few which have led to the measurement of one small-separation binary star and of two stellar angular diameters. We conclude with a brief outlook on further possible instrumental developments and an estimate of the scientific return.  In particular, we find that the LO technique can detect sources down to K$\approx 9$\,mag with SNR=1 on the DOT telescope. Angular diameters larger than $\approx 1$\,milliarcsecond (mas) could be measured with SNR above 10, or K$\approx 6$\,mag. These numbers are only an indication and will depend strongly on observing conditions such as lunar phase and rate of lunar limb motion.  Based on statistics alone, there are several thousands LO events observable in principle with the given telescope and instrument every year.
\end{abstract}


\section{Introduction}
\noindent Recently, the 3.6-m Devasthal Optical Telescope (DOT) 
started operations as the largest facility of its kind in India
\citep{2018BSRSL..87...29K}. It is located at 2450\,m elevation,
longitude of 79.7 E, latitude of 29.4 N,
and managed by the Aryabhatta Research Institute of Observational Sciences (ARIES; Nainital).
The telescope has a 3.6-m active-control primary mirror,
Ritchey-Chretien optics, an alt-azimuth mount 
and a corrected science field of view (FoV) of 30$\arcmin$ at the Cassegrain focus. It is equipped with various Imager and spectrographs operating
from the visual to the near-infrared (NIR) range 
\citep{2018JAI.....750003B,2018BSRSL..87...58O, 2019arXiv190205857O, 2022JApA...43...27K}.
The site is operated for eight months which are not affected by the monsoon rainfall, i.e., October to May, in two observational cycles.
The median seeing at Devasthal is around $1\farcs1$ (at ground level in optical bands), although best values can approach $0\farcs6$ at the 3.6-m DOT height \citep{2001BASI...29...39S}.

In this paper, we focus on the instrumental requirements and
scientific aspects relative to the observation of Lunar Occultation (LO)
events at DOT in the NIR. 
{ The LO technique consists in recording and analyzing the
diffraction pattern generated when the lunar limb moves over
a background source, and it is especially attractive since it
achieves angular resolution far exceeding the diffraction
limit of the telescope. Given the typical angular rate of the limb,
it generally requires time resolutions of few ms.} 
A strong tradition already 
started in India at the { 1.2-m Mount Abu telescope}
three decades ago \citep{1993BASI...21..499C} with a single-pixel
InSb detector and later continued with
a NIR focal plane array (FPA) detector
\citep[e.g.][]{2010MNRAS.408.1006C}. More recently,
observations have continued at ARIES using a
frame transfer Andor iXon EMCCD on the 
 1.3-m telescope also located at Devasthal
\citep[e.g.][]{{2017MNRAS.464..231R},
{2020MNRAS.498.2263R}}. This is however the first time that
a considerably larger telescope is employed for fast photometric
observations { in India.  }
A  mention of two initial LO events recorded with the TIRCAM2 { (TIFR Near Infrared Imaging Camera-II,
the only instrument on 3.6m DOT which is capable in doing millisecond sampling)} instrument 
at the 3.6-m DOT  was given in \citet{2020MNRAS.498.2263R}. In the present paper
we provide full details of this operational mode in Sect.~\ref{tircam2},
and present the complete commissioning results.
We will briefly
describe the method and the associated data analysis in
Sect.~\ref{LOmethod}.  A { total of 12}
LO events have been successfully recorded
at DOT using TIRCAM2 to date, of which one has led to the
measurement { of two binary stars}
 and two resolved stellar diameter.
We will discuss these results in
Sect.~\ref{LOresults}, while in 
Sect.~\ref{outlook} we will outline how we intend
to further improve the instrumental aspects in order to
achieve higher temporal resolution, and the scientific
strategy for this specialized mode of observation at DOT.

\begin{figure}[h]
\begin{center}
\includegraphics[width=0.75\textwidth, angle=0]{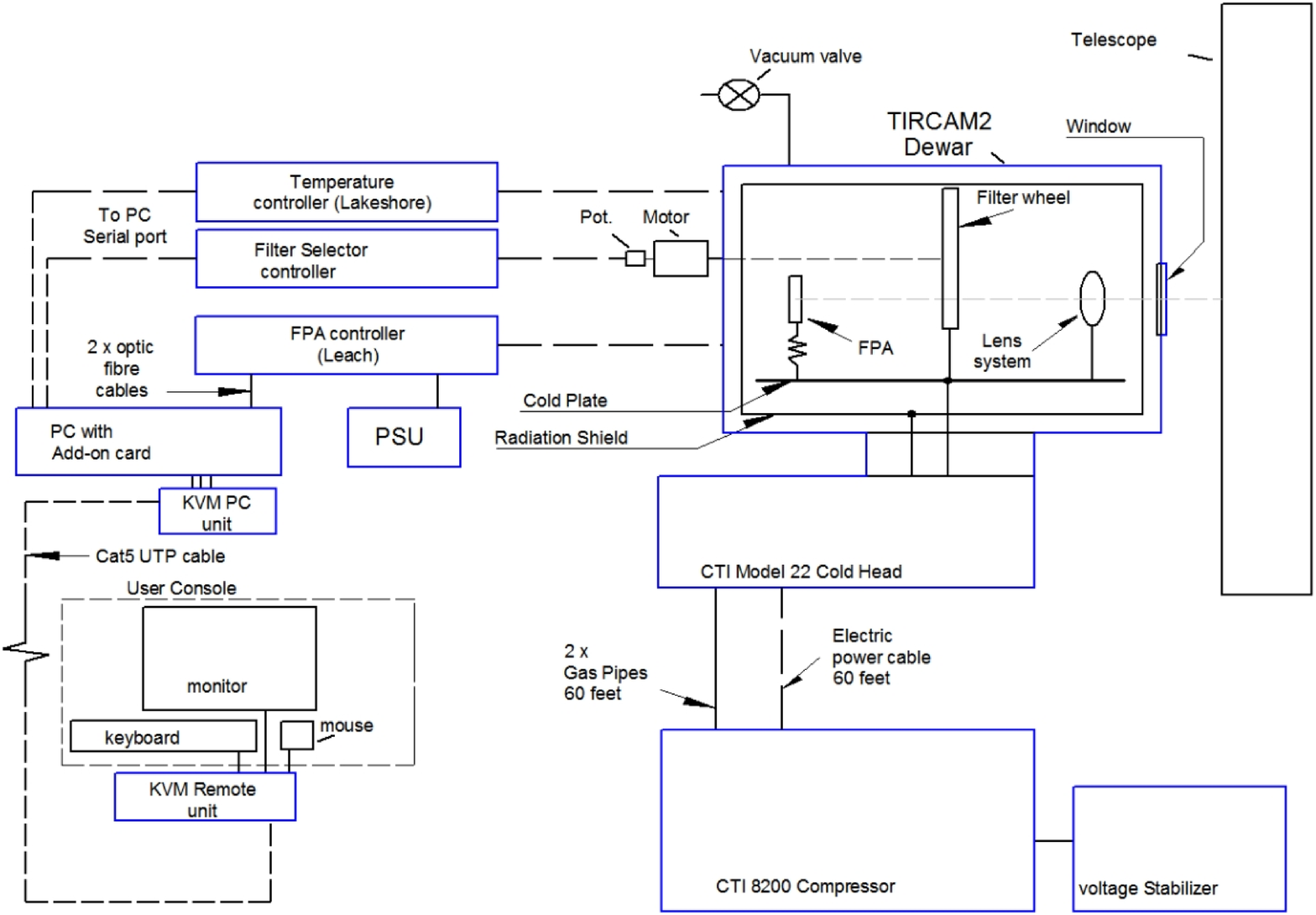} %100 percent
\end{center}
	\caption{ A schematic picture showing different components of TIRCAM2 which are mounted on the telescope (Dewar, Temperature/Filter/FPA controllers, cold head, PC etc), located on the telescope floor (compressor, voltage stabilizer) and located at the observing floor (remote access PC).}
\label{fig:tircam2}
\end{figure}

\subsection{103~Tau}\label{sec:103tau}

\section{TIRCAM2 and its Adaptation for Fast Photometry}\label{tircam2}

TIRCAM2 is a closed--cycle cooled Imager
that has been developed by the Infrared Astronomy Group at Tata Institute of
Fundamental Research for observations in the NIR range \citep{2012BASI...40..531N}.
TIRCAM2 is sensitive between 1 and 5\,$\mu$m and includes both the standard NIR
broad-band filters J, H, K, as well as several narrow-band filters, namely
K$_{\rm cont}$, Brackett-Gamma, Polycyclic Aromatic Hydrocarbon (PAH) and
L-narrow. 
 TIRCAM2 uses a $512\times512$
InSb Aladdin III Quadrant focal plane array,
which corresponds to 
$86\farcs5 \times 86\farcs5$ FoV on 3.6m DOT with a plate scale of $0\farcs169$/pixel. 
It is cooled to an operating temperature
of 35~K by a closed cycle Helium cryo-cooler.
{ A schematic picture of TIRCAM2 showing its different components is presented in Fig.~\ref{fig:tircam2}. 
More details on TIRCAM2 can be found in \citet{2022JApA...43...16G} and \citet{2018JAI.....750003B}.}
TIRCAM2 is currently the only NIR imaging camera in India which can observe up to L band in the NIR. 
{ Therefore, this camera could be a good complementary instrument to observe the 
bright nbL-band sources that are saturated in the Spitzer-Infrared Array Camera (IRAC) Ch1-band ([3.6] . 7.92 mag) 
and the WISE W1-band ([3.4] . 8.1 mag).  Sources with strong polycyclic aromatic hydrocarbon (PAH) 
emission at 3.3 micron are also observable with TIRCAM2 \citep{2018JAI.....750003B}.
In our initial LO commissioning observations reported here, we have used
only the broad-band K filter but any of the other available filters could be used too,
depending on the trade-off between sensitivity and sky background for the given
science case. }
The dark current measured is $\approx 12$~e$^-$/s and the readout noise $\approx 30$~e$^-$.
The median gain of the detector is $\approx 10$~e$^-$/ADU.

Due to the elevated sky background in the NIR, typical exposure times  per frame
with TIRCAM2
for conventional imaging are already considerably shorter than
for visual range instrumentation and of order $\approx 10$\,s.
In fact, individual exposures of the full focal plane array (FPA)
can be captured
in as short as 256\,ms. However, this is far too slow for our LO application
which requires individual exposures in the ms range.
Thus, we developed a fast photometry mode, in which the data
are captured
in $32\times32$ pixels sub-array, corresponding to
about $5\farcs4 \times 5\farcs4$ on the sky.
	As the full width at half-maximum of the stellar images  at the 3.6-m DOT is 
	typically $\sim 0\farcs7$ in the NIR,
this is sufficient for  photometry including a reasonable sampling of
the surrounding sky background. { The sub-array
 is   positioned around the target by providing its x \& y FPA coordinates to the TIRCAM2 software.}
 
    The read-out software  of TIRCAM2 comprises three individual programs residing in the computer,  
     in the PCI board DSP (Digital Signal Processor), and in the controller DSP, respectively. 
    To achieve a  faster readout, the sub-array mode required modifications in both the
    computer JAVA code and
    the controller DSP assembly code. 
   With the current version of the software,  the sampling time is $\sim$10 ms for sub-array mode of TIRCAM2.
	After the sub-array is programmed and initialized in the
    controller, it captures the required frames and the data are stored in a single FITS file on the computer.
   {
   In case of the capture box (32$\times$32 pixels) being at the center of FPA, 
   after the FPA is reset, it integrates light for $\sim$1.7 ms until capture box digitization is started. 
   Digitization continues for another $\sim$3.2 ms and then it continues skipping the remaining rows 
   for another $\sim$1.7 ms. The dead time between each frame is around 3.2 ms. 
   So, the total exposure time for the capture box is $\sim$4.9 ms and the total time between the consecutive 
   frames is $\sim$9.8 ms. As it takes $\sim$0.007 ms to skip a row of FPA, the exposure time depends on 
   where the star is located on FPA and can range from 3.2 ms to 6.6 ms. 
   It is possible to reduce the integration time by placing the capture box close to the start row of the FPA. This was not done for the present commissioning work.
   }
    For LO observation, we generally take 4000 frames { requiring 
    $\sim 40$\,s detector illumination,
    and about 2 min in total including overheads.  }

\section{The Lunar Occultation Technique}\label{LOmethod}
LO represented the method of choice to achieve 
milli-arcsecond (mas) resolution mainly in the 1970-90's.
{
We mention briefly that the
technique consists in recording the Fresnel diffraction pattern generated
when the lunar limb moves over a distant background source. The limb 
rate is typically $\approx 0.7$\,m/ms, the event lasts a small fraction of a second,
and detection rates of few ms are required to adequately sample the fringes.
The principles of data reduction were first formulated by
\citet{1970AJ.....75..963N}, while more details and advanced procedures
can be found e.g. in \citet{1996AJ....112.2786R}.
} 

Albeit significantly limited in the choice of targets by 
the orbit of the Moon and by being fixed-time events, LO offer several advantages 
including a very efficient use of telescope time, a fast and simple data reduction 
coupled with the ability to derive model-independent results, and an angular 
resolution which is rather independent of seeing and telescope diameter. 

Recently, 
more sophisticated methods like long-baseline interferometry 
and to some extent adaptive-optics and speckle interferometry  
are now available for similar studies at selected large facilities, but
they
are typically more time consuming and limited by the need of suitable
reference stars.
LO still occupy an attractive niche and offer a favorable combination of 
angular resolution and sensitivity especially at relatively small telescopes where 
other methods are not an option.

Following a large program of routine LO observations at the ESO VLT in the 
NIR which yielded over 1,000 recorded events about a decade ago with 
a corresponding large number of new angular diameters and binary stars
\citep[][and references therein]{2014AJ....147...57R}, 
several observatories have now implemented this technique. Examples include the 
Russian 6-m telescope, and several other facilities such as the 1.3-m telescope
also at Devasthal, the 1.2-m and 1.8-m telescopes at Asiago in Italy, 
the 2.4-m Thai National telescope, the 1.2-m MAO telescope in Trebur, Germany,
and the 3.5-m Telescope Nazionale Galileo (TNG) in La Palma, Spain. 

LO are essentially a 1-D technique, along the direction of the lunar limb
motion. Therefore, only the combined observation from different sites offers
the opportunity to obtain a 2-D view \citep[see e.g.][]{2017MNRAS.464..231R}.
Coordinated observations also provide the opportunity for multi-wavelength
coverage. In this respect, we stress that TIRCAM2 at DOT is ideally
positioned to complement similar observations from East Asia to Europe, and
that it is the only facility at present offering NIR coverage.

{
We remark that while LO observations are possible at any wavelength and there
are examples in the literature spanning from ultraviolet to mid-infrared,
the NIR range is often considered an ideal range. This is due to the combination
of various non-linear effects: the sky background (mainly scattering of the reflected solar
light from the Moon), the thermal background from the
lunar surface, the fringe speed decrease with wavelength (which translates to better fringe 
sampling for a fixed detector cycle), the spectral energy distribution of the occulted source
which is often rising towards longer wavelengths for evolved stars and object with dust envelopes.
}

LO data reduction is considerably less intensive than all other
high angular resolution techniques both in terms of data volumes
and in required computing power. It is also unique in the sense
of being rather insensitive to seeing conditions and without the
need for a reference star. Two different approaches are possible.
Firstly,
a model-dependent least-square method that
 includes the estimation of several parameters, including
the lunar limb rate which translates to the position angle of the
1-D scan, and 
the angular diameter of the occulted source. It is described e.g.
in \citet{1992A&A...265..535R}, while 
\citet{1996AJ....112.2786R} describe a method to put an upper
limit on the  diameter of unresolved sources depending on the noise in the data.
With a completely different approach based on an iterative
deconvolution technique,
\citet{1989A&A...226..366R} introduced a method (dubbed composite algorithm) to achieve
a model-independent
reconstruction of the brightness profile in the
maximum-likelihood sense.
Examples of the use of these methods are provided in
Sect.~\ref{LOresults}.

\section{Commissioning and First Results}\label{LOresults}

We report here the initial commissioning observations of the
fast photometric mode of TIRCAM2 at the 3.6-m DOT. They were all
carried out using a broad-band K filter centered at 2.2 $\mu$m.
The list of observations is provided in Table~\ref{aba:tbl1},
 where the columns are largely self-explanatory. The predicted time is listed
rounded down to the nearest minute, while El and Ill are the
 Moon's elevation above the horizon and the illuminated fraction. 
 A negative or positive Ill value denotes reappearance or disappearance, respectively. 
 The entries in the next columns are from the  SIMBAD database, 
 and SNR is the signal-to-noise ratio for the best possible fit
 using, depending on the case, models of a point-source, of a resolved stellar diameter,
 or of a binary star. Some comments are given under Notes, where UR stands
 for unresolved.
 {
 We clarify that in case of good SNR, it is possible to put an upper limit on
 the angular size even if it remains unresolved, as is the case for SAO~77838.
 }
%} 
 
 \begin{wstable}[h]
\caption{List of LO events observed with TIRCAM2 at DOT.}
\begin{tabular}{lccrrcclrc} \toprule
\multicolumn{1}{c}{Source}	&	
\multicolumn{1}{c}{Date}	&	
\multicolumn{1}{c}{Time}	&	
\multicolumn{1}{c}{El}	&
\multicolumn{1}{c}{Ill}	&
\multicolumn{1}{c}{V}	&	
\multicolumn{1}{c}{K}	&	
\multicolumn{1}{c}{Sp} &		

\multicolumn{1}{c}{SNR}	&	
\multicolumn{1}{c}{Notes} \\ 
\colrule
TYC639-399-1	&	07-Oct-17	&	15 47	&	25$\degr$	&	$-95$\%	&	10.8	&	6.5	&	M2 III	&	\multicolumn{1}{c}{-}	&		\\
TYC642-458-1	&	07-Oct-17	&	18 15	&	56$\degr$	&	$-94$\%	&	10.7	&	6.8	&		&	\multicolumn{1}{c}{-}	&		\\
Mu Cet	&	07-Oct-17	&	20 04	&	71$\degr$	&	$-94$\%	&	4.1$^a$	&	3.5	&	A9IIIp	&	\multicolumn{1}{c}{-}	&	($^1$)	\\
IRC +20156	&	18-May-18	&	14 32	&	23$\degr$	&	$+13$\%	&	8.0	&	2.6	&		&	30.4	&	UR$^2$	\\
	SAO 98770	&	21-May-18	&	20 04	&	22$\degr$	&	$+45$\%	&	9.4	&	7.8	&	F8	&	18.1	&	binary$^3$ ($177\pm0.4$ mas)	\\
SAO 165154	&	12-Nov-21	&	17 27	&	21$\degr$	&	$+63$\%	&	9.0	&	6.2	&	K1III	&	\multicolumn{1}{c}{-}	&		\\
BD+00149	&	15-Nov-21	&	20 32	&	20$\degr$	&	$+89$\%	&	7.0	&	4.2	&	G8III	&	34.1	& UR \\
	IRC +10032	&	15-Dec-21	&	14 42	&	68$\degr$	&	$+90$\%	&	7.6	&	2.6	&	M	&	71.9	&	diameter ($2.22\pm0.25$ mas)	\\
HD 284115	&	14-Jan-22	&	12 45	&	45$\degr$	&	$+90$\%	&	8.1	&	4.2	&	K2	&	17.8	&	UR	\\
HD 32380	&	14-Jan-22	&	13 17	&	51$\degr$	&	$+90$\%	&	8.3	&	5.2	&	K0III	&	\multicolumn{1}{c}{-}	&		\\
	103 Tau	&	14-Jan-22	&	16 34	&	83$\degr$	&	$+91$\%	&	5.5	&	5.3	&	B2V	&	12.55	&	binary ($16.8\pm0.4$ mas)	\\
	IRC +20101	&	14-Jan-22	&	18 57	&	54$\degr$	&	$+91$\%	&	8.9	&	2.7	&	M0	&	39.2	&	diameter ($2.39\pm0.28$ mas)	\\
SAO 77800	&	15-Jan-22	&	14 59	&	63$\degr$	&	$+95$\%	&	6.6	&	4.5	&	K0III	&	12.9	&	UR	\\
	SAO 77838	&	15-Jan-22	&	16 36	&	83$\degr$	&	$+96$\%	&	9.1	&	3.8	&	K7	&	66.5	&	upper limit ($1.45\pm0.65$ mas)	\\
IRC +30138	&	15-Jan-22	&	22 03	&	26$\degr$	&	$+96$\%	&	7.6	&	2.6	&	M5Ib	&	\multicolumn{1}{c}{-}	&		\\
IRAS 04269+2252	&	09-Mar-22	&	14 41	&	54$\degr$	&	$+43$\%	&	11.3$^b$	&	4.7	&	IR Source	&	4.0	&	UR	\\
IRAS 04278+2253	&	09-Mar-22	&	15 20	&	46$\degr$	&	$+43$\%	&	15.0$^a$	&	5.9	&	F1, YSO	&	9.0	&	UR	\\
IRC +20084	&	09-Mar-22	&	15 59	&	37$\degr$	&	$+43$\%	&	7.1	&	2.9	&	K5III	&	16.4	&	UR	\\
\\
\botrule
\end{tabular}
\begin{tablenotes}
\item[1] Observed simultaneously at the 1.3-m telescope
\item[2] 16\,ms sampling
\item[3] First near-IR measurement. Discussed in detail in \citet{2020MNRAS.498.2263R}. 32\,ms sampling
\item[a, b] R and G magnitudes, respectively \\
Events with "-" in the SNR column were not detected, for reasons ranging
from sensitivity to pointing issues to timing.
\end{tablenotes}
\label{aba:tbl1}
\end{wstable}
 
It can be seen that a first run on October 7, 2017 did not produce
usable data also given the difficulty of blind pointing due 
to the events being reappearances,  but it allowed us to fine tune the electronics parameters and
read-out process. 
{ Since the window size of 32$\times$32 pixels on 3.6-m DOT corresponds to only $\sim5\farcs3\times5\farcs3$ on sky, 
the blind pointing is not advised in the case of the re-appearance events 
(the telescope usually have more $\sim$2 arcsec rms pointing errors). What we can do is to calculate 
the pointing offsets from a pre-identified nearby star and apply those to the target star during the event. 
The same procedure was adopted while carrying our re-appearances events from the 
1.3-m Devasthal Fast Optical Telescope (DFOT) \citep{2020MNRAS.498.2263R}.}
We also note that some of the events in Table~\ref{aba:tbl1}
occurred with a very high lunar phase.
This resulted in a higher than usual
 background due to proximity of sunlight reflected off the lunar surface,
 and more photon noise in the data as a consequence.
We expect that under average conditions the SNR
for similar magnitudes might be better than reported in Table~\ref{aba:tbl1}. 
We emphasize that the few selected results
that we discuss in the following are affected by 
relatively large uncertainties, as is to be expected given the
slower than ideal time sampling, and in
general by the commissioning nature of the observations. They are
mainly intended to verify the feasibility of LO with TIRCAM2, and
to encourage future users of this method.  We are also striving to improve
the SNR performance in the near future, as discussed in Sect.~\ref{outlook}.

\subsection{IRC~+10032}\label{sec:irc10032}
\begin{figure}[h]
\begin{center}
\includegraphics[scale=0.7, angle=-90]{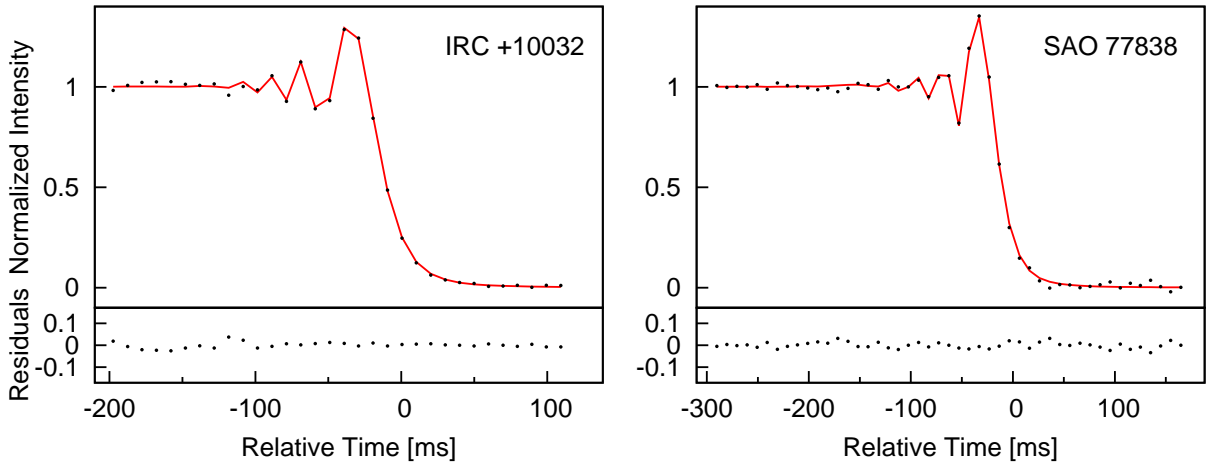} %100 percent
\end{center}
\caption{{\it Left:} LO data normalized in intensity for IRC~+10032 (dots) 
and best fit by a model with a stellar diameter of 2.2\,mas (line)
are shown in the top panel. {\it Right:} the same for SAO~77838,
fitted by a point-source model. Note the higher fringe contrast for
the unresolved star case. The two bottom panels show the respective
fit residuals. Details are given in the text.
}
\label{fig:irc10032}
\end{figure}

Our data for IRC~+10032 are shown on the left side of
Fig.~\ref{fig:irc10032}, together with a best fit
by a uniform disk (UD) angular diameter of
$2.22\pm0.25$\,mas.
Interestingly, the first high angular resolution measurement
of this M giant was also a LO in the K-band reported by \citet{2000MNRAS.317..687T},
who found it unresolved with an upper limit of 2\,mas.
Recently, \citet{2022MNRAS.510...82R} accurately measured the UD diameter
of IRC~+10032 (HD~17973) to be $1.93\pm0.015$\,mas in the L band.
Our measurement is roughly consistent with this value.

\subsection{103~Tau}\label{sec:103tau}
\begin{figure}[h]
\begin{center}
\includegraphics[scale=0.6, angle=-90]{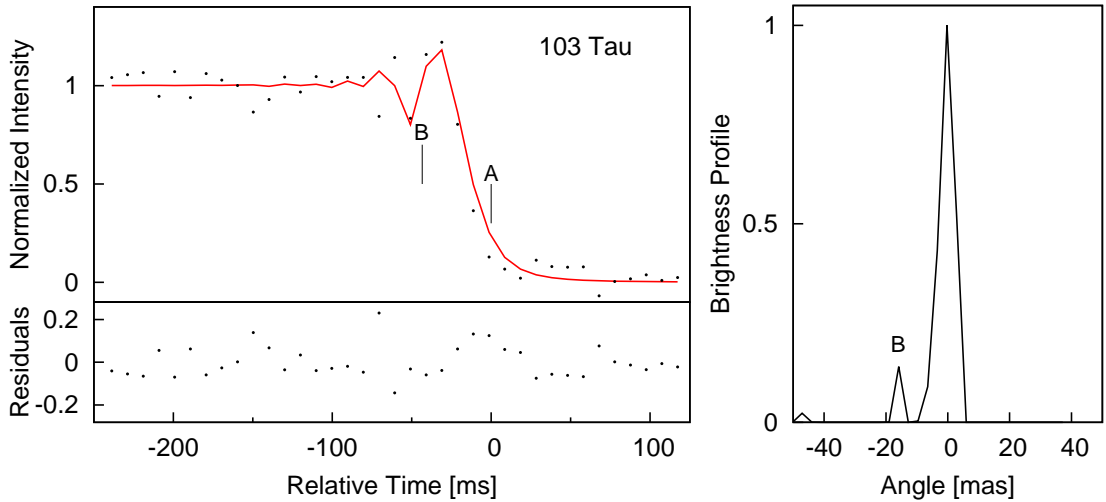} %100 percent
\end{center}
\caption{{\it Left:} LO data normalized in intensity for 103~Tau (dots) 
and best fit by a binary model (line)
are shown in the top panel, while the bottom panel shows the fit residuals.
The position in time of the components is shown, with the secondary being
occulted about 43\,ms 
before the primary. {\it Right:} a model-independent brightness
profile reconstruction, showing the binary nature of 103~Tau. Details are given in the text.
}
\label{fig:103tau}
\end{figure}

The data and fit for 103~Tau are shown in Fig.~\ref{fig:103tau}, together
with a model-independent reconstruction of the brightness profile obtained
with the CAL method mention in Sect.~\ref{LOmethod}. 
{
The reduced contrast of the main fringe is highly suggestive
of the presence of a companion, which is clearly revealed
in the model-independent analysis, see Fig.~\ref{fig:103tau}.
} 
From the
binary fit we find a projected separation of 
$16.8\pm 0.4$\,mas along position angle PA=$165\degr$, and a brightness difference from the primary of $2.31\pm0.04$\,mag.

The companion to 103~Tau had already been detected from a number
of investigations using speckle interferometry,
e.g. among the most recent ones
\citet{2015AJ....150..151H},
\citet{2020MNRAS.495..806G},
This is however the first time that it is detected in the NIR,
and our flux ratio determination of 1:8.3($\pm0.3$) well agrees
with the spectral classification of primary and secondary
being both B dwarfs with a
few sub-classes difference,
see e.g. \citet{2016AstL...42..598T}
using spectroscopic investigations.

\subsection{IRC~+20101}\label{sec:irc20101}

For this source, we find that a fit by a resolved uniform stellar
diameter of $2.39\pm0.28$\,mas is significantly better then the
fit by a point-like source (SNR$=39.2$ and 27.0 for the two
cases, respectively).
 The relatively large error is
mainly due to the  slow time sampling, but the value is 
approximately consistent
with the  estimate of the angular diameter for this M0 giant which is 1.9\,mas 
using the empirical calibration by \citet{1999PASP..111.1515V}.

There are no previous direct determinations of the angular diameter of {IRC~+20101}
in the literature,
and no observations by high angular resolution methods including no other LO
observations.

\subsection{SAO~77838}\label{sec:sao77838}

The data and best fit by a point-source for SAO~77838 are shown on the right side of
Fig.~\ref{fig:irc10032}. The SNR ratio for this data set 
is  one of the highest
obtained during commissioning, and we have used the method referenced 
in Sect.~\ref{LOmethod} to put an upper limit on the angular size of
SAO~77838 of $1.45\pm0.65$\,mas. 
This can be satisfactorily compared with
an empirical estimate of 1.1\,mas for the angular diameter of this
K7 giant,  located at $\approx 138$\,pc, 
 using the calibration by \citet{1999PASP..111.1515V}.
It reinforces 
the validity of using TIRCAM2 and the LO method
to measure angular diameters in the mas range.

As was the case for IRC~+20101, also for SAO~77838 there are no previous
LO observations reported and no relevant literature entries with direct
measurements.

\section{Planned Developments and Outlook}\label{outlook}

\begin{figure}[h]
\begin{center}
\includegraphics[scale=0.6, angle=-90]{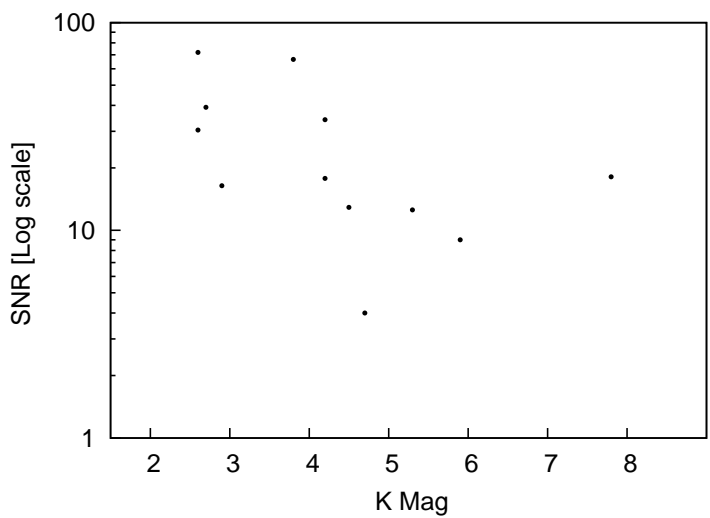} %100 percent
\end{center}
\caption{Plot of the SNR of the LO best fit as a function of the
K magnitude, using the applicable data from Table~\ref{aba:tbl1}. 
The point close to K=8\,mag (SAO~98770) stands out from the main trend
because it was recorded at much slower rate than the rest.
Discussion are given in the text.
}
\label{fig:ksnr}
\end{figure}

Using those data from Table~\ref{aba:tbl1} for which a fit could be made, 
we plot in Fig.~\ref{fig:ksnr} the SNR as a function of the K magnitude.
The scatter is significant, as expected given that LO events occur under a
large range of sky background (strongly dependent in turn on the illumination
fraction and the distance to the terminator) and other parameters.
Nevertheless, a trend is clearly visible already in these preliminary
commissioning data, with SAO~98770 being an outlier in virtue of
the sampling and integration time being 4 times longer than for the rest.
We can { thus} conclude that under the standard sub-array configuration
($32\times32$ pixels at $\sim 10$\,ms) the fast photometry mode of
TIRCAM2 can be sensitive to about K=9\,mag, 
{
to detect the occultation of a single unresolved sources, i.e. with SNR$\approx 1$.
As a rough guideline, the detection of  1:1 binary would require a SNR$\gtrapprox 3$, while
angular diameter determination would require a SNR$\gtrapprox 10$, or K$\lessapprox  6$\,mag.
}

This opens the possibility of using LO to investigate angular diameters
and binaries with targets available essentially any night that the Moon is up and 
the phase 2 days away from full Moon.
To put this in context, we have used the 2MASS catalog truncated to
the limit K$\le 8.5$\,mag to compute occultations for the whole year
of 2023 visible from the Devasthal site. Within acceptable limits
of lunar phase and elevation, a total of 18460 LO events would be
observable in principle over the year, of which 8766 disappearances.
{
As each LO event takes less than an half hour of telescope time (including overheads, initial settings etc), this program is ideal to fill the gaps in the telescope schedule and to utilize the less demanded bright moon period. However, due to non availability of an automated observational sequencer at the telescope, this program demands availability of dedicated and trained manpower at the site. In order to efficiently conduct this LO program from the telescope, development of both the observing sequencer and data reduction pipeline is required.}

The current main limitation is the sampling time, which at 10\,ms is
2 or 3 times longer than optimal for LO events 
{
\citep[see e.g.][]{1996AJ....112.2786R}.
}
We have been considering solutions, at both the
software and the hardware level.
 A software approach with the current FPA controller may be limited due to the limitations of the current TIRCAM2 hardware.
 A more promising approach would be the 
 installation of a new faster commercial controller, or the in-house design of a new one. 
 These possibilities are currently under exploration.

\section{Conclusions}
The TIRCAM2 instrument at the DOT 3.6-m telescope in Devasthal, northern India,
is an Imager originally designed to cover
the J to L near-infrared bands ($\approx 1-4~\mu$m) with an
InSb Aladdin $512\times512$ detector. The minimum full frame
read-out time is 256\,ms.
Having in mind the requirements
for LO observations, we have developed a fast
photometry mode which makes use of a $32\times32$ sub-array
with read-out time just under 10\,ms.

This mode has been commissioned by observing 18 LO events,
of which 12 were successful in recording the characteristic diffraction
patterns. Among them, we could measure two stellar angular diameters
and two binary stars. The rest of the data were consistent with
unresolved point sources. The sensitivity is established to
be close to K$\sim 9$\,mag. This in turn translates to several
thousands of LO events observable in principle from the Devasthal site each year.

The observation procedures and the data flow, resulting in 3-D standard
FITS cubes, have been thoroughly tested and the mode is now 
publicly offered to interested users. Improvements are being considered
to bring the sampling time down to very few milliseconds.

\section*{Acknowledgments}
We dedicate this paper to the memory of the late Prof. Anil K. Pandey, who 
led the establishment of a LO program at ARIES.
This work has made use of the SIMBAD data base, operated at CDS,
Strasbourg, France. It is a pleasure to thank the technical team for the
3.6m DOT, as well as the members of the IR astronomy group at TIFR,
for their support during the observing runs. SS acknowledges the support of the Department of Science and Technology, Government of
India, under project No. DST/INT/Thai/P-15/2019. DKO and MBN
acknowledge the support of the Department of Atomic Energy, Government of India, under Project Identification No.  RTI 4002.

\end{document}